\documentclass[twocolumn,prb,aps,showpacs,superscriptaddress]{revtex4-1}
\usepackage{graphicx}% Include figure files
\usepackage[dvipdfm]{hyperref}
\usepackage{amsmath}
\usepackage{epsfig}
\usepackage{color}

\begin{document}
\title{NMR detection of dynamical processes in antiferroelectric nanoclusters during the order-disorder transition in NH$_4$H$_2$AsO$_4$}

\author{Riqiang Fu}
\email{rfu@magnet.fsu.edu} 
\affiliation{National High Magnetic Field Laboratory, Florida State University, 1800 East Paul Dirac Drive, Tallahassee, Florida 32310, USA}

\author{Ozge Gunaydin-Sen} 
\affiliation{National High Magnetic Field Laboratory and Department of Chemistry and Biochemistry, Florida State University, 1800 East Paul Dirac Drive, Tallahassee, Florida 32310, USA}
\affiliation{Department of Chemistry and Biochemistry, Lamar University, Beaumont, Texas 77710, USA}

\author{Irinel Chiorescu} 
\affiliation{Physics Department and the National High Magnetic Field Laboratory, Florida State University, 1800 East Paul Dirac Drive, Tallahassee, Florida 32310, USA}

\author{Naresh S. Dalal}
\email{dalal@chem.fsu.edu} 
\affiliation{National High Magnetic Field Laboratory and Department of Chemistry and Biochemistry, Florida State University, 1800 East Paul Dirac Drive, Tallahassee, Florida 32310, USA}

\date{Phys. Rev. B Rapid Commun.  \textbf{91}, 140102(R) (2015)}%

\begin{abstract}
We study the dynamics of inorganic antiferroelectric nanoclusters formed during an order-disorder transition and demonstrate the coexistence of the two phases in a region of $2-3$~K around the transition temperature $T_N\sim$ 215~K. Single crystals of NH$_4$H$_2$AsO$_4$, a model hydrogen-bonded compound, show an antiferroelectric-paraelectric transition studied by means of highly sensitive magic angle spinning $^{15}$N NMR at 21.1 T. The phase co-existence is demonstrated by a double-peak structure of the chemical shift. Two-dimensional chemical exchange spectroscopy and spin-lattice relaxation time ($T_1$) measurements show that the clusters are dynamic with sizes $\sim$50~nm and lifetimes approaching seconds as $T\rightarrow T_N$. Their occupancy increases rapidly to fill the crystal volume below $T_N$. This study provides evidence for the commonality of the phase transitions in systems with electric properties and provides an improved spectroscopic method for such studies. \end{abstract}


\maketitle

An atomic-scale view of order-disorder transitions can significantly improve the understanding and theoretical modeling of interactions in the time domain, at the nanoscale level. Generally speaking, hydrogen-bonded compounds, such as the KH$_2$PO$_4$ (KDP)-family\cite{1Lines2001}, are considered to be significant for ferroelectric (FE) transitions from several points of view: (a) They serve as models of the phenomenon of hydrogen-bonding that is of fundamental importance in the physics and chemistry of ice and water \cite{2Krumhansl1990} as well as the structure and function of biological molecules such as DNA\cite{3Balasubramanian1998}; (b) the KDP family finds extensive applications in electro-optic devices because of their ferroelectric behavior \cite{1Lines2001}; and (c) despite recent advances in the theoretical modeling of FE and related order-disorder phase transitions via computer simulations \cite{4Ogita1969,5ScheiderPRL1973} and analytical solutions\cite{6Bussmann1997,7Bussmann1998,8Bussmann2009,9Bishop2010}, a general model of the transition dynamics is not available \cite{10McMahon1990,11Schmidt1987}. We present a study of the antiferroelectric (AFE) phase transition in an inorganic compound NH$_4$H$_2$AsO$_4$ (henceforth ADA) \cite{13Blinc1969, 14Adrianenssens1975, 15Adrianenssens1976,16Blinc1986} tailored to target the nanoscale and dynamical aspects of such a transition, which is the missing piece of many theoretical approaches. One long-standing unanswered query is that while evidence of pretransitional cluster formation has been seen in the case of the FE transitions (in KDP systems\cite{13Blinc1969, 14Adrianenssens1975,15Adrianenssens1976,16Blinc1986,12Lasave2007} KH$_2$AsO$_4$ and RbH$_2$AsO$_4$ using $^{75}$As NMR and nuclear quadrupole resonance), similar measurements have failed to detect pretransitional clusters during the AFE transitions\cite{15Adrianenssens1976}. Such a failure casts some doubt on the universality of the dynamics accompanying the ferroelectric phenomenon in general, since computer models predict a similar picture of the phase transition dynamics in any lattice that undergoes an order-disorder phase transition \cite{4Ogita1969,5ScheiderPRL1973,6Bussmann1997,7Bussmann1998,8Bussmann2009,9Bishop2010}.

The results of the above-mentioned NMR experiments\cite{13Blinc1969,14Adrianenssens1975, 15Adrianenssens1976,16Blinc1986}, together with the theoretical predictions\cite{4Ogita1969,5ScheiderPRL1973,6Bussmann1997,7Bussmann1998,8Bussmann2009,9Bishop2010}, suggest that in the ADA-type AFE lattices, the pretransition clusters must be at the nanoscale, thus requiring increased spectral resolution and sensitivity. The ratio $\eta=\nu_0/\delta\nu$, where $\nu_0$ is the signal frequency and $\delta\nu$ the linewidth is an accepted figure of merit for spectral resolution. For ADA, the previous \cite{14Adrianenssens1975} $\eta$ was 1T/(1mT)$\sim 10^3$. The recent development of magic angle spinning\cite{Stejskal1994} (MAS) NMR techniques at 21.1 T at the National High Magnetic Field Laboratory (NHMFL), gives $\eta\sim$(91$\times$10$^6$ Hz)/5 Hz$\sim$2$\times 10^7$ for the $^{15}$N nucleus, constituting a four order-of-magnitude enhancement of the spectral resolution\cite{18Fu2007}. This technique enables the detection of the low-symmetry AFE clusters (co)existing in the high-symmetry [paraelectric (PE)] phase of ADA. These pretransitional AFE clusters are dynamic and we present experimental evidence that atoms are moving between the clustered region and the rest of the lattice. The results provide support for theoretical models studying such dynamic atomic exchange and cluster formation when approaching an order-disorder phase transition\cite{4Ogita1969,5ScheiderPRL1973,6Bussmann1997, 7Bussmann1998,8Bussmann2009,9Bishop2010}. 

We employed a $^{15}$N-enriched ADA crystal with a size of 1x1x1 mm$^3$ and $\sim9\%$ enrichment, prepared as reported earlier \cite{18Fu2007}. ADA is a colorless transparent crystal that at room temperature crystallizes in a body-centered-tetragonal unit cell \cite{10McMahon1990}, space group $I\bar{4}2d$, with the unique ($c$-) axis cell length = 7.709~\AA, and $a=b=$7.674~\AA. The structure changes to an orthorhombic ($P$2$_1$2$_1$2$_1$) phase at temperatures below the AFE phase transition, measured to be $\sim$215 K. It becomes antiferroelectric, with the H's localized asymmetrically along one of the O's [1,10,11]. As a spectral reference, we use powdered Valine. The spin-lattice relaxation time $T_1$ is measured by enhancing the saturation of the nuclei via a standard cross polarization (CP) protocol \cite{Stejskal1994}. In addition, the MAS technique allows us to obtain isotropic resonance signals by spinning the sample around an axis tilted by an angle of 54.7$^\circ$ away from the applied magnetic field\cite{Stejskal1994}. At this magic angle, the inhomogeneous dipolar broadening is averaged out by the fast spinning. The NMR measurements were made on the ultrawide bore 900 MHz (21.1 T) NMR spectrometer at NHMFL with a 5~kHz spinning speed and temperature stability of 0.1~K. In this setup, the $^{15}$N and  $^1$H Larmor frequencies are $\omega_N/2\pi=90.91$~MHz and $\omega_H/2\pi=897.19$~MHz, respectively.
 
\begin{figure}
	\includegraphics[width=3.25 in, , bb=5 62 766 560]{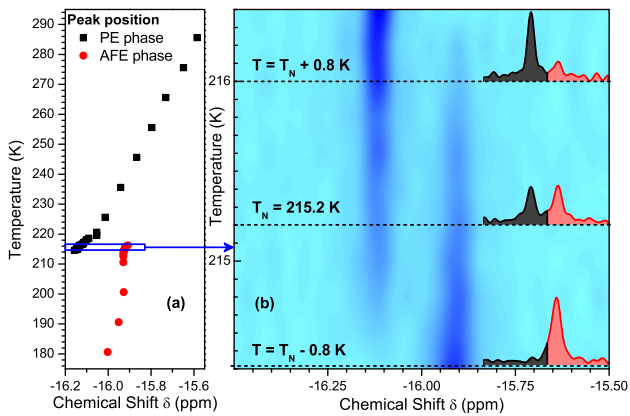}
	\caption{(Color online) (a) $^{15}$N isotropic chemical shift of ADA, showing an abrupt change  at $T_N=$215.2~K as a function of temperature. Data points indicate the maxima of $^{15}$N CPMAS spectra, such as shown in (b). (b) Contour plot of CPMAS spectra (arbitrary units) in the vicinity of $T_N$, with darker areas indicating the location of maxima or high signals. The insets show three spectra; the vertical axes are of same range for all three and the horizontal axis is identical to the one of the contour plot. $T_N$ was defined as the temperature at which the PE (dark gray) and AFE (light red) peaks have equal integrated intensities.}
	\label{fig1}
\end{figure}

The isotropic chemical shift $\delta_{iso}= (\omega -\omega_N)/\omega_N$ is a parameter characterizing magnetic shielding effects of an atom due to cyclotronic motion of local charges in a fixed magnetic field, with $\omega$ being the actual, measured resonance frequency of the $^{15}$N nuclei in a given environment. This parameter provides detailed information on the local chemical environment at the atomic level. The temperature dependence of the $^{15}$N $\delta_{iso}$ is shown in Fig.~\ref{fig1}(a). The $^{15}$N signal from PE is shown by black squares and that of AFE by red dots. One observes that $\delta_{iso}$ exhibits a continuous behavior in both high and low temperature phases but with different slopes, separated by a sudden change around the transition temperature $T_N$. In the narrow transition region one observes the coexistence of two maxima in the cross polarization and magic angle spinning (CPMAS) spectra as detailed in Fig.~\ref{fig1}(b). Within about 2-3~K of $T_N$ the crystal lattice consists of both phases. The PE and AFE phases were clearly resolved owing to the exceptionally high spectral resolution \cite{18Fu2007} by MAS at 21.1~T (peak widths of about 0.1 ppm). Note that at any given temperature, the NMR peak areas represent the relative abundance of the PE and AFE fractions in the crystal lattice. The insets show such spectra at $T_N=215.2$~K, where the areas of the two peaks are about equal, while below and above the transition temperature the areas are different. It is evident the transfer of intensity between the AFE peak (light red) and the PE peak (dark gray) as the temperature is lowered through $T_N$.  This shows that one phase is consumed by the other, reversibly, with very little temperature hysteresis if the temperature variation is slow, about 0.1 K per 10 min, as $T\rightarrow T_N$.

\begin{figure}
\includegraphics[width=3.25 in, bb=19 27 729 553]{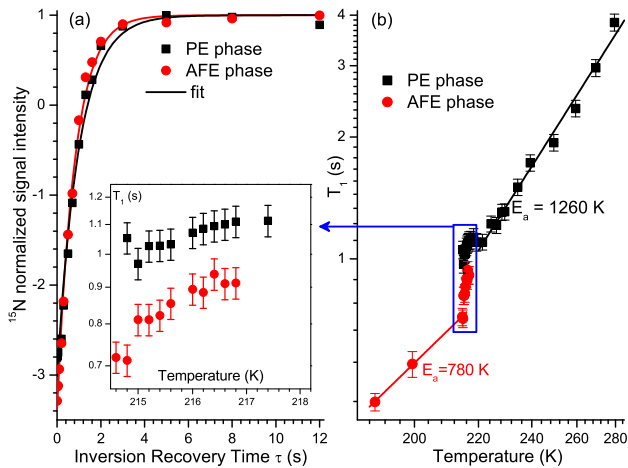}
\caption{(Color online) (a) Normalized $^{15}$N magnetization $M_z^{CP}$ of the PE (squares) and AFE (dots) phases vs the inversion recovery time $\tau$ at $T_N$. Single exponential fittings (lines) yield $T_1$ of 1.03 and 0.82~s for the PE (black) and AFE (red) phases, respectively. The inset is a zoom of data in (b) as indicated by the blue rectangle. (b) Temperature dependence of $^{15}$N $T_1$ in ADA shown in logarithmic scale for the vertical axis and reciprocal scale ($1/T$) for the temperature axis, to put in evidence the linear behavior [see Eqs.~(\ref{eq_T1}) and (\ref{tauc})].  }
\label{fig2}
\end{figure}

The high resolution of the peaks allows us to measure the spin-lattice relaxation time $T_1$ of the two phases outside but also inside the coexisting temperature regime. A standard inversion recovery method was used, and the signal recovery to the equilibrium following an inversion pulse was recorded as a function of recovery time $\tau$. Two examples are shown in Fig.~\ref{fig2}(a), one for each phase at $T_N$. Briefly after CP, the enhanced $^{15}$N polarization is flipped to the $-Z$ axis (that is, antiparallel to the applied field). The $^{15}$N magnetization $M_z^{CP}$ then relaxes back to its equilibrium (i.e., $+Z$ axis) along the longitudinal direction. This recovery process can be characterized by $T_1$ as follows, 
\begin{equation}
M_z^{CP}(\tau)=1-\alpha\exp(-\tau/T_1)
\end{equation}
where $\alpha$ is the cross polarization efficiency. Theoretically $\alpha$ could be as large as 10 ($\sim \gamma_H/\gamma_N$, the ratio of $^1$H and $^{15}$N gyromagnetic ratios, respectively), but the experimental setups and relaxation properties could greatly affect the CP efficiency\cite{19Fu2004}. In our experiments, $\alpha$ was about 4-4.5. As shown in Fig.~\ref{fig2}(a), a single exponential fitting yields different $T_1$ values. Interestingly, for the peak representing the PE phase, $T_1$ was found to be 1.03$\pm$0.05~s, while for the AFE peak it was significantly shorter, 0.82$\pm$0.04~s, both measured at the same temperature ($T_N$).

\begin{figure}
	\includegraphics[width=3.25 in, bb=55 35 664 525]{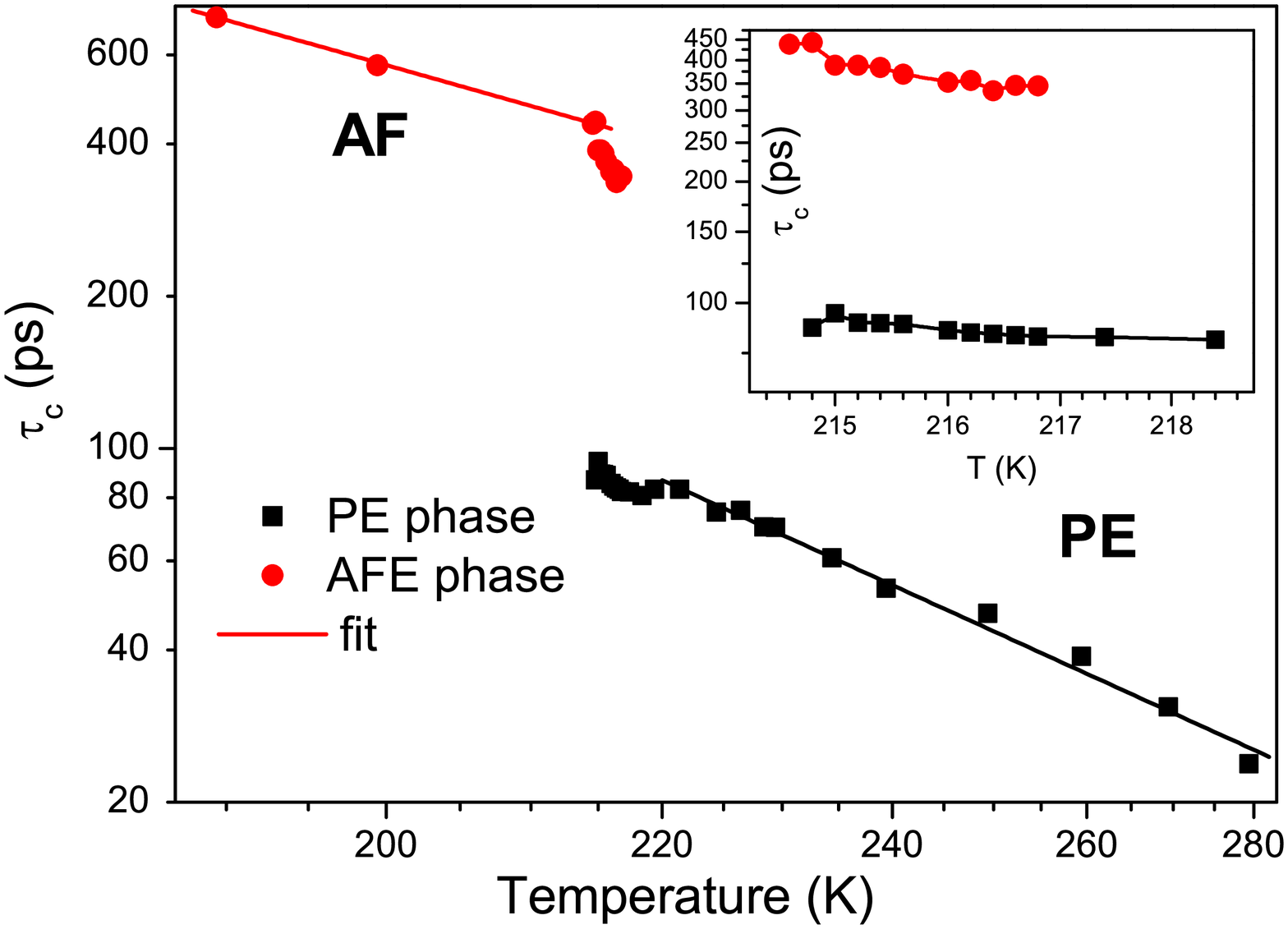}
	\caption{(Color online) Correlation time as a function of temperature, for the AFE (dots) and PE (squares) phases shown in logarithmic scale for the vertical axis and reciprocal scale ($1/T$) for the temperature axis. The lines show Arrhenius fits done using Eq.~\ref{tauc}. The inset shows the same data points (no fits) in the same representation, inside the order-disorder transition (error bars are smaller than dot sizes).}
	\label{fig3}
\end{figure}

The nuclear relaxation time $T_1$ is an important tool in studying phases and their transition. The temperature dependence of the $^{15}$N $T_1$ from 295~K down to 180~K is shown in Fig.~\ref{fig2}(b). A clear discontinuity occurs in the $T_1$ data near $T_N\sim$215~K, implying that the two phases have distinctly different motional correlation times. The $T_1$ data of Fig.~\ref{fig2}b were analyzed using the Bloembergen-Purcell-Pound (BPP) model\cite{20BPP1947}:
\begin{eqnarray}\label{eq_T1}
    \frac{1}{T_1}&=&\frac{C\tau_c}{1+(\omega_H-\omega_N)^2\tau_c^2} +\frac{3C\tau_c}{1+\omega_N^2\tau_c^2}+\\
&&\frac{6C\tau_c}{1+(\omega_H+\omega_N)^2\tau_c^2}.\nonumber
\end{eqnarray}
Here, $\tau_c$ is the correlation time describing the dynamic process and $C=\frac{1}{10}(\frac{\mu_0}{4\pi}\frac{\gamma_N\gamma_H\hbar}{r^3_{NH}})^2\approx4.86\times10^8$~s$^{-2}$ is the dipole-dipole relaxation constant, which is associated with the nitrogen-proton distance (N-H bond distance) $r_{NH}=1.031 \AA$ taken from Ref.~\onlinecite{Gunaydin-Sen2006}. Since $T_1$ is of the order of a second, this results in $\tau_c\sim 1$ns $\sim \omega_{N,H}^{-1}$ and consequently Eq.~(\ref{eq_T1}) can be reduced to $T_1^{-1}\sim10C\tau_c$. For thermally activated diffusion, $\tau_c$ is assumed to follow the Arrhenius behavior,
\begin{equation}\label{tauc}
\tau_c=\tau_0\exp\frac{E_a}{kT}
\end{equation}
where  $\tau_0$ is the inverse of a vibrational mode related to the phase transition, $E_a$ the activation energy,  $k$ the Boltzmann constant, and $T$ the temperature. The experimental $T_1^{-1}$ data in Fig.~\ref{fig2}(b) were fitted to the above equations, with $E_a$ and $\tau_0$ being treated as variables. The slopes of the linear fittings $\ln T_1$ vs $T^{-1}$, as indicated by the solid lines in Fig.~\ref{fig2}(b), give different $E_a$ for the PE and AFE phases, of 1260~K (or 10.2 kJ/mol) and 780~K  (or 6.3 kJ/mol), respectively. Similarly, the Arrhenius fits given in Fig.~\ref{fig3} give $\tau_0$ equal to 0.3 and 11.6~ps for the PE and AFE phases, respectively (and the same values for the $E_a$ parameter). The difference in $E_a$ correlates well with the fact that the transition induces a significant change of the crystalline structure. We ascribe the increase in $\tau_0$ as due to the effective ``soft" mode leading to the AFE transition becoming frozen below $T_N$. This indicates that the slowing down of the NH$_4^+$ torsional motion triggers the phase transition, likely due to the formation of new N-H$\cdots$O bonding~\cite{11Schmidt1987,12Lasave2007}. Moreover, as shown below, we can extract microscopic information on the size and mass exchange of the ordered phase nanoclusters.

\begin{figure}
\includegraphics[width=3.25 in]{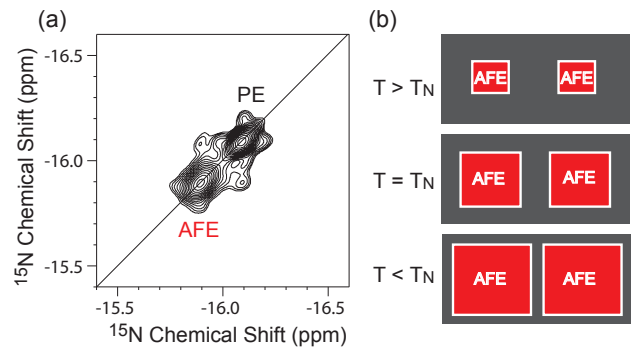}
\caption{(Color online) (a) 2D $^{15}$N exchange spectrum of ADA at 215.2~K with a mixing time of 1~s. The presence of the cross peaks indicates a dynamic exchange of the nuclei from the AFE clusters to the PE clusters. (b) Schematic of the AFE clusters in the PE clustered lattice in going from the PE phase (black) to the AFE phase (red).}
\label{fig4}
\end{figure}

As mentioned earlier on, a major goal of this study was to measure the spatial dimensions of the pretransitional clusters just above $T_N$, their lifetimes and whether there is any exchange between the atomic units on the periphery of the clustered domain and the PE domains. In NMR, such chemical exchange experiments are carried out by introducing a \emph{mixing time} in between two $\pi/2$ pulses, preceded by an evolution time $t_1$, and followed by a detection time $t_2$. This pulse sequence\cite{Levitt2008} allows us to record the measured signals during $t_{1,2}$ (the $^{15}$N chemical shifts) when the two phases evolve separately. During the mixing time, however, the chemical shifts start to crosstalk if their corresponding spins (or atoms) are involved in a dynamic process, which results in off-diagonal cross peaks after a two-dimensional (2D) Fourier transform. Figure~\ref{fig4}(a) shows a typical $^{15}$N 2D chemical exchange spectrum of ADA at 215~K ($\sim T_N$) recorded with a mixing time of 1.0~s. The two axes represent the Fourier transforms of the signals measured during the times $t_1$ (vertical) and $t_2$ (horizontal). At this temperature, within 0.1~K of $T_N$, the coexistence of both the PE and AFE regions is evidenced by the diagonal peaks (at -16.1 and -15.9 ppm), so marked.  The 2D spectrum also clearly shows the presence of cross peaks, albeit small, $\sim$5\% as compared to the diagonal peaks. The existence of the cross peaks implies that the PE and AFE clusters are in a dynamic equilibrium such that there is a diffusion\cite{Levitt2008} of molecules between the two phases, and moreover that a temperature gradient across the small sample is not the cause of the cluster formation. This latter was also checked by carrying out measurements on a crystal half the size of the previous one, and no significant change in the extent of the peak coexistence temperature was observed. We interpret the appearance of the cross peaks as showing that some of the PE clusters are converted into the AFE domains and vice versa although such intercluster exchanges are on the slow time scale of the order of the mixing time, $\sim$1 s, at $T_N$. Obviously the cluster sizes and lifetimes should decrease as the crystal temperature is raised well above $T_N$. Figure~\ref{fig4}(b) shows the schematic of the AFE clusters in the PE clustered lattice in going from the PE phase to the AFE phase assessed from the relative intensities of the diagonal and cross peaks in Fig.~\ref{fig4}(a). The cross peaks yield the relative abundance of the interfacial volume between the PE and AFE clusters, i.e., the molecules that are exchanging between the PE and AFE domains. These are represented by the white boxes in a 2D representation shown in Fig.~\ref{fig4}. The diagonal peaks yield the relative abundance of the PE (black) and AFE (red) domains. 

We then carried out a simple calculation of an estimate of the length scale of the clusters as they start to appear on the NMR timescale and resolution. Since the AFE crystal structure is an orthorhombic ($P$2$_1$2$_1$2$_1$) phase, it is reasonable to assume that the AFE clusters are of cubic shape with a side length $l$, giving a total surface area of $6l^2$ and a cluster volume of $l^3$. If the thickness of the cluster surface is assumed to be about the size of the unit cell, i.e., $\sim$ 770~pm, the ratio of the boundary between the PE and AFE clusters and the AFE cluster volume becomes 770pm$\times 6l^2/l^3=4620$ pm/$l$. At $T_N$, the PE and AFE clusters give about the same signal intensity (diagonal peaks), so that the chances of the signals from the boundary area going to the PE and AFE clusters are about the same. Thus, the ratio of the observed cross peak intensity versus the diagonal peak is proportional to 2310~pm$/l$. From Fig.~\ref{fig4}(a), the cross peak intensities are about 5\% of the diagonal peak intensities, yielding an average cluster size of $\sim$45 $\pm$ 5~nm. Therefore, as the PE lattice approaches its transition temperature, it develops finite size clusters that represent the symmetry of the low-temperature phase (i.e., AFE phase), and these AFE clusters increase in size and numbers to fill the entire lattice as the temperature goes down below 215.2~K, as illustrated schematically in Fig.~\ref{fig4}(b).

Concluding, we detail the structural and dynamic aspects of this antiferroelectric transition and give evidence for the co-existence of two phases during such an order-disorder transition. This was achieved using the exceptionally high spectral resolution provided by MAS NMR at 21.1 T. The quantitative estimate of the AFE clusters size is $\sim$45~nm and their lifetimes are $\sim$1 s at $T_N$+0.1~K. These quantitative data should serve for the development of precise models of AFE phase transitions. The reported NMR technique should also open a window on other phase transitions, especially for all nonconducting and nonmagnetic solids for which high-resolution solid NMR is a sensitive structural and molecular dynamics probe.

\acknowledgements
We are grateful for financial support from the User Collaboration Grant Program (UCGP) at the National High Magnetic Field Laboratory (NHMFL) which is supported by the NSF Cooperative Agreement No. DMR-1157490, the State of Florida, and the U.S. Department of Energy. A portion of this work was supported by NSF Grant No. DMR-1206267.

\end{document}